\documentclass[reprint, amsmath,amssymb, aps, prb]{revtex4-2}

\usepackage[english]{babel}
\usepackage{graphicx}
\usepackage{dcolumn}
\usepackage{dsfont}
\usepackage{bm}
\usepackage{xcolor}

\DeclareUnicodeCharacter{2212}{-}
\DeclareUnicodeCharacter{03B4}{$\delta$}
\usepackage{hyperref}

\newcommand{\ZZ}{\mathrm{Z\kern-.3em\raise-0.5ex\hbox{Z}}}

\newcommand{\I}{\mathrm{i}}

\newcommand{\ket}[1]{\left| #1 \right\rangle}

\newcommand{\brakett}[2]{\left\langle #1 \middle| #2 \right\rangle}
\newcommand{\brakettt}[3]{\left\langle #1 \middle| #2 \middle| #3 \right\rangle}

\begin{document}

\title{Polarization dependency in Resonant Inelastic X-Ray Scattering}
\author{Michelangelo Tagliavini}
\author{Fabian Wenzel}
\author{Maurits W. Haverkort}

\affiliation{Institute for Theoretical Physics, Heidelberg University, Philosophenweg 19, 69120 Heidelberg, Germany}

\date{\today}

\begin{abstract}

Resonant Inelastic X-Ray Scattering (RIXS) is a well-established tool for probing excitations in a wide range of materials. The measured spectra strongly depend on the scattering geometry, via its influence on the polarization of the incoming and outgoing light. By employing a tensor representation of the 4-point response function that governs the RIXS intensity, we disentangle the experimental geometry from the intrinsic material properties. In dipole-dipole RIXS processes and low-symmetry crystals, up to 81 linearly independent fundamental spectra can be measured as a function of light polarization. However, for crystals or molecules with symmetry, the number of independent fundamental spectra that define the RIXS tensor is significantly reduced. 

This work presents a systematic framework for determining the number of fundamental spectra and expressing the RIXS tensor in terms of these fundamental components. Given a specific experimental geometry, the measured spectrum can be represented as a linear combination of these fundamental spectra. To validate our approach, we performed calculations for different point group symmetries, both with and without an applied magnetic field. Within the same framework, we derived expressions for powder spectra in momentum-independent processes and spectra obtained using Bragg spectrometers. This formalism provides a valuable toolkit for optimizing experiment planning, data interpretation, and RIXS simulation.

\end{abstract}

\maketitle


\section{\label{sec:Intro}Introduction}

Resonant Inelastic X-Ray Scattering (RIXS) is a spectroscopic technique in which an X-ray photon is resonantly absorbed and then coherently emitted from a sample. The electric dipole probability of this process is measured as a function of excitation energy ($\hbar \omega_{\text{in}}$), emission energy ($\hbar \omega_{\text{out}}$), momentum transfer ($\hbar \mathbf{q}$), and the polarization of the incoming and outgoing light ($\boldsymbol{\hat{\epsilon}}_{\mathrm{in}}$ and $\boldsymbol{\hat{\epsilon}}_{\mathrm{out}}$). 

Advances in radiation source brilliance and improvements in experimental resolution have enabled RIXS to flourish over the past two decades \cite{dallera_soft_1996,chuang_high-resolution_2005, ghiringhelli_saxes_2006,alonso-mori_multi-crystal_2012,shvydko_merixnext_2013,szlachetko_situ_2013,boots_optimizing_2013,chiuzbaian_design_2014, dvorak_towards_2016,zimina_cat-actnew_2017,miyawaki_compact_2017,moretti_sala_high-energy-resolution_2018,abela_swissfel_2019,ablett_galaxies_2019,jaeschke_synchrotron_2020,nowak_versatile_2020,gretarsson_irixs_2020,rovezzi_texs_2020,weinhardt_x-spec_2021,kokkonen_upgrade_2021,singh_development_2021,glatzel_five-analyzer_2021,bauer_mev_2022,zhou_i21_2022,schlappa_heisenberg-rixs_2025}. RIXS became a powerful tool for investigating the local electronic state including its spin state \cite{hayashi_selective_2005,pirngruber_presence_2006}, valence orbital occupations \cite{wu_advances_2020,pidchenko_synthesis_2020,schacherl_resonant_2025}, as well as low-energy charge \cite{ghiringhelli_long-range_2012,hepting_three-dimensional_2018,kang_resolving_2019,lin_doping_2020} , spin \cite{braicovich_dispersion_2009, glawion_two-spinon_2011, le_tacon_intense_2011, zhou_persistent_2013}, orbital \cite{ghiringhelli_nio_2005, ghiringhelli_resonant_2006, schlappa_spinorbital_2012}, and lattice \cite{yavas_observation_2010,moser_electron-phonon_2015,dashwood_probing_2021} excitations in a broad range of materials. The use of X-ray energies provides bulk sensitivity and element selectivity \cite{kotani_resonant_2001, groot_core_2008}. The RIXS scattering cross-section is theoretically described by the Kramers-Heisenberg equation \cite{kramers_uber_1925,schulke_electron_2007}. Although the light-matter interaction and the Kramers-Heisenberg formalism are well understood, calculating RIXS spectra in real materials remains far from trivial \cite{luo_scattering_1993, ament_ultrashort_2007,haverkort_theory_2010, kas_real-space_2011, lu_nonperturbative_2017,tsvelik_resonant_2019,hariki_lda_2020,thomas_theory_2025}. For small molecules and clusters the active space can be efficiently restricted and numerical solutions can be obtained \cite{glatzel_electronic_2004,guo_molecular_2016,polly_relativistic_2021}, albeit in many cases at substantial computational costs. For extended systems an effective operator theory exists, \cite{haverkort_theory_2010} substantially reducing the calculation complexity and providing a tractable understanding of the RIXS cross-section. The number of effective operators needed for a realistic experiment can however be quite large \cite{lu_nonperturbative_2017} hindering an intuitive understanding of RIXS in, for example, metals. Upon resonant absorption of an X-ray photon, an excited state with a core hole is created. These states involve strong Coulomb interactions and, for core shells with angular momentum, significant core spin-orbit coupling. This leads to the formation of local excitons or resonances with atomic-like multiplet structures \cite{thole_3d_1985,de_groot_2p_1990,haverkort_multiplet_2012}. While atomic multiplets involving both core and valence open shells are theoretically well-defined, their interpretation can be arduous. Developing an intuitive understanding of RIXS spectra has therefore proven to be challenging. In this work, we focus on the polarization dependence of RIXS and demonstrate how tensor formulation, a concept widely used for the conductivity tensor in (nonlinear) optics \cite{armstrong_interactions_1962,maker_study_1965,duboisset_nonlinear_2025}, or x-ray absorption spectroscopy \cite{haverkort_symmetry_2010}, provides a simplified framework to understand the polarization dependence in RIXS experiments.

Resonant excitations in the X-ray range are predominantly governed by electric dipole photon-matter interactions. The wavelengths of X-rays can be comparable to the spatial extent of the valence orbitals involved, thus higher-order multipole transitions might be expected to play a role. However, due to the participation of core states in the excitation process, which spatial extent is reduced, the same multipole transitions are mostly suppressed. Multipole transitions in the X-ray range can still be observed, albeit generally with lower intensity than dipole transitions \cite{bartolome_quadrupolar_1999,wende_quadrupolar_2002,hayashi_quadrupole_2004,glatzel_electronic_2013}. While X-ray absorption spectroscopy (XAS) and RIXS theories for multipole transitions have been extensively discussed in the literature \cite{luo_scattering_1993, carra_high_1995, ament_resonant_2011,juhin_angular_2014}, in this work we focus exclusively on electric dipole transitions.

RIXS can be employed to probe local excitations when both the intermediate state, reached via resonant excitation, and the final state, detected in the emission process, involve a core hole. For example, in actinide materials, a $3d$ electron can be excited into the $5f$ valence shell, followed by the decay of a $4f$ electron into the created $3d$ core hole \cite{kvashnina_resonant_2017,vitova_role_2017}. This allows the investigation of shallow core-excited states using tender X-rays. The resulting spectroscopy technique shares similarities with XAS and core-level X-ray photoelectron spectroscopy (XPS) but offers enhanced sensitivity. Since core orbitals are highly localized and do not disperse, RIXS spectra in this limit are, to first order, independent of the transferred momentum $\mathbf{q}$. Instead, momentum transfer is accommodated by the entire lattice or by phonon and vibrational modes, which have energies comparable to or smaller than the Auger-Meitner and fluorescence linewidths of the core-excited states and thus play a minor role in understanding the so called core-to-core RIXS spectra.

RIXS can also be used to probe dispersive low-energy valence excitations. In this case, the X-ray photon excites a core electron into the valence states, and subsequently, a valence electron decays back into the core hole, leaving behind a low-energy excitation. This channel is referred core-to-valence RIXS. This allows to measure dispersing excitations such as magnons, phonons, orbitons, and charge excitations. Low-energy charge excitations observed with RIXS can be compared to those studied in optical \cite{hunault_direct_2018} or Raman spectroscopy. At zero momentum transfer ($\mathbf{q}=0$), these spectroscopies probe similar final states as RIXS but follow different selection rules. Similarly, dispersing magnetic excitations probed with RIXS can be compared to inelastic neutron scattering (INS) \cite{braicovich_dispersion_2009}. However, while the selection rules for exciting low-energy magnetic excitations in INS are relatively simple, the effective low-energy operators in RIXS are significantly more complex~\cite{haverkort_theory_2010}. This complexity arises because RIXS can excite multiple low-energy excitations, including double, triple, and quadruple magnons, as well as charge excitations \cite{elnaggar_magnetic_2023}. Disentangling these contributions requires careful analysis of the resonance and polarization dependence of RIXS spectra.

This work offers a practical and systematic toolkit for computing and characterizing the polarization dependence of RIXS spectra. For example one can relate the recently reported circular dichroism in RIXS\cite{biniskos_systematic_2025, jost_chiral_2025, takegami_circular_2025, furo_theory_2025} to symmetry related fundamental spectra. The method is applicable to any crystal or molecular point group and fully leverages symmetry-based selection rules, here used in the special case of momentum-independent scattering. We build on previous studies that have explored polarization effects in RIXS. For instance, the angular-averaged scattering of a powder sample, as presented in Section~\ref{sec:AvvPowder}, reproduces the results obtained from spherical tensor coupling considerations by Juhin \textit{et al.}~\cite{juhin_angular_2014}.  For core-to-valence RIXS, the same formalism can be extended employing space-group based arguments. With the modification that for finite momentum scattering at scattering vector $\mathbf{q}$ we need to consider the little group, i.e. the subgroup of the crystallographic point group that leaves $\mathbf{q}$ invariant modulo reciprocal lattice vectors, instead of the full point-group of the crystallographic site. 

The paper begins by defining the experimental geometry, particularly the polarization vectors of the incident and scattered light in crystal coordinates. The next section introduces the RIXS tensor, followed by a discussion on symmetry considerations in Section~\ref{sec:Symmetry}, which identifies the tensor elements that vanish due to symmetry constraints. Sections~\ref{sec:NumExample1} and~\ref{sec:NumExample2} provide numerical examples illustrating the formalism. Finally, we discuss three common experimental setups: (i) measurements without polarization analysis of the scattered photons (Section~\ref{sec:AvvEpsOut}), (ii) measurements on powdered samples (Section~\ref{sec:AvvPowder}), and (iii) the effects of Bragg reflection in analyzer crystals or gratings used to select the energy of the scattered photons (Section~\ref{sec:Analyzer}).

\section{\label{sec:Scatteringgeometry}Scattering geometry}

\begin{figure}[htb]
    \includegraphics[width=0.65\linewidth]{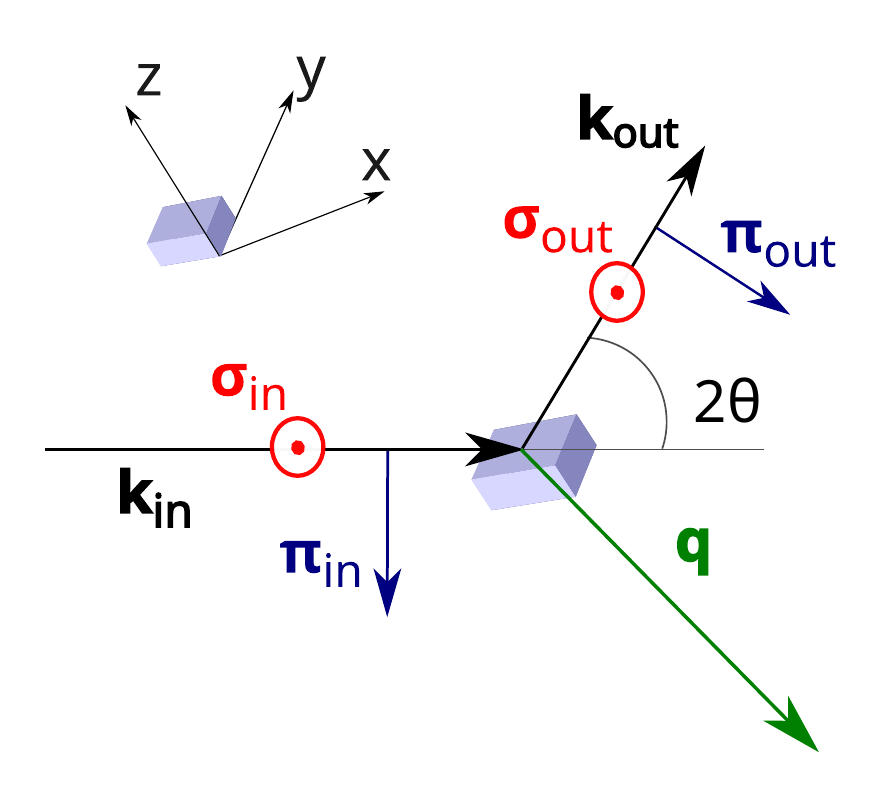}
    \caption{\label{fig:FigScatteringGeometrySample} 
    Schematic representation of the RIXS experimental setup. The photon momentum is denoted as $\hbar\mathbf{k}$, while the transferred momentum to the sample is given by $\hbar\mathbf{q}$. The polarization basis vectors are defined as $\boldsymbol{\sigma}$, which is perpendicular to the scattering plane (vertical), and $\boldsymbol{\pi}$, which lies within the scattering plane (horizontal). The coordinate system is determined by the sample orientation, independent of the scattering plane.}
\end{figure}

Figure~\ref{fig:FigScatteringGeometrySample} illustrates the experimental configuration and the parameters necessary to uniquely describe a RIXS experiment. Incident light with momentum $\hbar \mathbf{k}_{\mathrm{in}}$ and polarization $\boldsymbol{\hat{\epsilon}}_{\mathrm{in}}$ scatters off the sample at an angle of $2\theta$, emerging with momentum $\hbar \mathbf{k}_{\mathrm{out}}$ and polarization $\boldsymbol{\hat{\epsilon}}_{\mathrm{out}}$. Here the hat indicates that the vector is normalized. The momentum transferred to the sample is given by $\hbar \mathbf{q}$, where
\begin{align}
    \mathbf{q} = \mathbf{k}_{\mathrm{in}} - \mathbf{k}_{\mathrm{out}}.
\end{align}

The vectors $\mathbf{k}_{\mathrm{in}}$ and $\mathbf{k}_{\mathrm{out}}$ define the scattering plane, which, in Figure~\ref{fig:FigScatteringGeometrySample}, corresponds to the plane of the paper. However, the sample orientation is not necessarily aligned with this plane. To describe the scattered intensity in a physically meaningful way, it is convenient to adopt the coordinate system of the sample, thereby directly linking the experiment to the intrinsic material properties.

Given $\mathbf{k}_{\mathrm{in}}$ and $\mathbf{k}_{\mathrm{out}}$ in crystal coordinates, we require a suitable method to represent the polarization in the same coordinate frame. The polarization vector $\boldsymbol{\hat{\epsilon}}$ must be perpendicular to the corresponding wave vector $\mathbf{k}$. To achieve this, it is common practice to define the basis vector $\boldsymbol{\hat{\sigma}}$ as perpendicular to the scattering plane and $\boldsymbol{\hat{\pi}}$ as perpendicular to both $\mathbf{k}$ and $\boldsymbol{\hat{\sigma}}$. These basis vectors are determined as follows:
\begin{align} \nonumber 
\boldsymbol{\hat{\sigma}}_{{\mathrm{in}}({\mathrm{out}})} &= \frac{\mathbf{k}_{\mathrm{in}} \times \mathbf{k}_{\mathrm{out}}}{|\mathbf{k}_{\mathrm{in}}| |\mathbf{k}_{\mathrm{out}}| \sin(2\theta)}, \\ 
\boldsymbol{\hat{\pi}}_{{\mathrm{in}}({\mathrm{out}})} &= \frac{\mathbf{k}_{{\mathrm{in}}({\mathrm{out}})}}{|\mathbf{k}_{{\mathrm{in}}({\mathrm{out}})}|} \times \boldsymbol{\hat{\sigma}}_{{\mathrm{in}}({\mathrm{out}})}.
\label{eq:polarizationdef}
\end{align}

For a general experiment, the polarization vectors $\boldsymbol{\hat{\epsilon}}_{\mathrm{in}}$ and $\boldsymbol{\hat{\epsilon}}_{\mathrm{out}}$ can be expressed in terms of the $\boldsymbol{\hat{\sigma}}$ and $\boldsymbol{\hat{\pi}}$ basis as:
\begin{align} \nonumber
\boldsymbol{\hat{\epsilon}}_{\mathrm{in}}(\alpha_{\mathrm{in}},\beta_{\mathrm{in}}) &= \cos(\alpha_{\mathrm{in}}) \boldsymbol{\hat{\pi}}_{\mathrm{in}} + \sin(\alpha_{\mathrm{in}}) e^{i \beta_{\mathrm{in}}} \boldsymbol{\hat{\sigma}}_{\mathrm{in}}, \\
\boldsymbol{\hat{\epsilon}}_{\mathrm{out}}(\alpha_{\mathrm{out}},\beta_{\mathrm{out}}) &= \cos(\alpha_{\mathrm{out}}) \boldsymbol{\hat{\pi}}_{\mathrm{out}} + \sin(\alpha_{\mathrm{out}}) e^{i \beta_{\mathrm{out}}} \boldsymbol{\hat{\sigma}}_{\mathrm{out}},
\label{eq:poldefinition}
\end{align}
where $\alpha_{\mathrm{in}(\mathrm{out})}, \beta_{\mathrm{in}(\mathrm{out})} \in [0,2\pi)$. These definitions express the polarization state of the light in terms of its $\boldsymbol{\hat{\sigma}}$ and $\boldsymbol{\hat{\pi}}$ components, which are referenced to the scattering plane and, via Eq.~\eqref{eq:polarizationdef}, can be related to crystal coordinates. Furthermore, in cases where the outgoing polarization is not analyzed, the corresponding measurement can be modeled by averaging (integrating) over $\alpha_{\mathrm{out}}$ and $\beta_{\mathrm{out}}$ (cf. Section~\ref{sec:AvvEpsOut}).

\section{\label{sec:TheRIXSTensor}The RIXS tensor}

The intensity of RIXS spectra within the dipole approximation depends on the energy ($\hbar \omega_{\mathrm{in}}$) and polarization ($\boldsymbol{\hat{\epsilon}}_{\mathrm{in}}$) of the photon used to excite the sample, the energy ($\hbar \omega_{\mathrm{out}}$) and polarization ($\boldsymbol{\hat{\epsilon}}_{\mathrm{out}}$) of the scattered photon, and the transferred momentum $\mathbf{q}$. Following the Kramers-Heisenberg equation and using Kubo's notation for response functions \cite{kubo_statistical-mechanical_1957} written in the frequency domain or Lehmann representation, the fraction of photons scattered into a solid angle $\partial\Omega$ and an energy transfer window from $\omega$ to $\omega + \partial \omega$ is proportional to
\begin{widetext}
\begin{align}
\label{eq:KramersHeisenberg}
\frac{\partial^2 \sigma}{\partial \Omega \partial \omega}  \propto  -\mathrm{Im}  \brakettt{\Psi_0} { T_{\mathbf{k}_{\mathrm{in}},\boldsymbol{\hat{\epsilon}}_{\mathrm{in}}}^{\dagger} G^{\dagger}(\omega_{\mathrm{in}}) T_{\mathbf{k}_{\mathrm{out}},\boldsymbol{\hat{\epsilon}}_{\mathrm{out}}}^{\phantom{\dagger}} G(\omega) T_{\mathbf{k}_{\mathrm{out}},\boldsymbol{\hat{\epsilon}}_{\mathrm{out}}}^{\dagger} G(\omega_{\mathrm{in}}) T_{\mathbf{k}_{\mathrm{in}},\boldsymbol{\hat{\epsilon}}_{\mathrm{in}}}^{\phantom{\dagger}}}{\Psi_0},
\end{align}
\end{widetext}
where $\omega = \omega_{\mathrm{in}}-\omega_{\mathrm{out}}$. $T_{\mathbf{k},\boldsymbol{\hat{\epsilon}}}$ is the operator describing the absorption of a photon with wave vector $\mathbf{k}$ and polarization $\boldsymbol{\hat{\epsilon}}$, $\ket{\Psi_0}$ is the ground state of the system, and $G(\omega) = 1/(\hbar \omega - H + E_0 + \I0^+)$ is the Green’s function describing the propagation of the system with Hamiltonian $H$ at an energy $\hbar \omega$ above the ground-state energy $E_0$.

As a first step, we simplify the Kramers-Heisenberg equation \eqref{eq:KramersHeisenberg} by approximating the dependence on $\mathbf{k}_{\mathrm{in}}$ and $\mathbf{k}_{\mathrm{out}}$ separately, reducing it to a single dependence on $\mathbf{q} = \mathbf{k}_{\mathrm{in}} - \mathbf{k}_{\mathrm{out}}$. Following Ref.~\cite{haverkort_theory_2010}, we rewrite the expression as
\begin{align}
\frac{\partial^2 \sigma}{\partial \Omega \partial \omega}  \propto  -\mathrm{Im} \brakettt{\Psi_0}{\left(R_{\mathbf{k}_{\mathrm{in}}\mathbf{k}_{\mathrm{out}}}^{\boldsymbol{\hat{\epsilon}}_{\mathrm{in}}\boldsymbol{\hat{\epsilon}}_{\mathrm{out}}}\right)^{\dagger} G(\omega) R_{\mathbf{k}_{\mathrm{in}}\mathbf{k}_{\mathrm{out}}}^{\boldsymbol{\hat{\epsilon}}_{\mathrm{in}}\boldsymbol{\hat{\epsilon}}_{\mathrm{out}}}}{\Psi_0},   
\end{align}
where $R_{\mathbf{k}_{\mathrm{in}}\mathbf{k}_{\mathrm{out}}}^{\boldsymbol{\hat{\epsilon}}_{\mathrm{in}}\boldsymbol{\hat{\epsilon}}_{\mathrm{out}}}$ is the RIXS transition operator describing the process in which a photon is scattered, thereby creating an excitation in the system:
\begin{align}
R_{\mathbf{k}_{\mathrm{in}}\mathbf{k}_{\mathrm{out}}}^{\boldsymbol{\hat{\epsilon}}_{\mathrm{in}}\boldsymbol{\hat{\epsilon}}_{\mathrm{out}}} = T_{\mathbf{k}_{\mathrm{out}},\boldsymbol{\hat{\epsilon}}_{\mathrm{out}}}^{\dagger} G(\omega_{\mathrm{in}} = c |\mathbf{k}_{\mathrm{in}}|) T_{\mathbf{k}_{\mathrm{in}},\boldsymbol{\hat{\epsilon}}_{\mathrm{in}}}^{\phantom{\dagger}}.
\end{align}
The transition operator $T_{\mathbf{k},\boldsymbol{\hat{\epsilon}}}$ can be written as a sum over operators acting on atomic sites with coordinates $\mathbf{R}_j$. Within the dipole approximation, the local transition operator $T_{j,\boldsymbol{\hat{\epsilon}}}$ acting on site $j$ is independent of the direction of $\mathbf{k}$, allowing us to write
\begin{align}
    T_{\mathbf{k},\boldsymbol{\hat{\epsilon}}} = \sum_j e^{\I \mathbf{k}\cdot \mathbf{R}_j} T_{j,\boldsymbol{\hat{\epsilon}}},
\end{align}
and
\begin{align}
\label{eq:RRIXSIntermediateStep}
R_{\mathbf{k}_{\mathrm{in}}\mathbf{k}_{\mathrm{out}}}^{\boldsymbol{\hat{\epsilon}}_{\mathrm{in}}\boldsymbol{\hat{\epsilon}}_{\mathrm{out}}} = \sum_{j',j} e^{\I \left(\mathbf{k}_{\mathrm{in}} \cdot \mathbf{R}_j - \mathbf{k}_{\mathrm{out}} \cdot \mathbf{R}_{j'}\right) }T_{j',\boldsymbol{\hat{\epsilon}}_{\mathrm{out}}}^{\dagger} G(c |\mathbf{k}_{\mathrm{in}}|) T_{j,\boldsymbol{\hat{\epsilon}}_{\mathrm{in}}}^{\phantom{\dagger}}.
\end{align}
The operator $T_{j,\boldsymbol{\hat{\epsilon}}}$ creates a core hole at site $j$, which we assume to remain immobile in the crystal. Consequently, the hole created at site $j$ must be annihilated at the same site ($j' = j$) in Eq.~\eqref{eq:RRIXSIntermediateStep}. Defining $\mathbf{q} = \mathbf{k}_{\mathrm{in}} - \mathbf{k}_{\mathrm{out}}$, we obtain
\begin{align} \nonumber
R_{\mathbf{k}_{\mathrm{in}}\mathbf{k}_{\mathrm{out}}}^{\boldsymbol{\hat{\epsilon}}_{\mathrm{in}}\boldsymbol{\hat{\epsilon}}_{\mathrm{out}}} &= \sum_j e^{\I \mathbf{q}\cdot \mathbf{R}_j}  T_{j,\boldsymbol{\hat{\epsilon}}_{\mathrm{out}}}^{\dagger} G(c |\mathbf{k}_{\mathrm{in}}|) T_{j,\boldsymbol{\hat{\epsilon}}_{\mathrm{in}}}^{\phantom{\dagger}} \\ \nonumber
&= \sum_j e^{\I \mathbf{q}\cdot \mathbf{R}_j} R_{\omega_{\mathrm{in}},j}^{\boldsymbol{\hat{\epsilon}}_{\mathrm{in}}\boldsymbol{\hat{\epsilon}}_{\mathrm{out}}} \\ 
&= R_{\omega_{\mathrm{in}},\mathbf{q}}^{\boldsymbol{\hat{\epsilon}}_{\mathrm{in}}\boldsymbol{\hat{\epsilon}}_{\mathrm{out}}}.
\label{eq:qDependentEffectiveR}
\end{align}
The double differential RIXS cross-section is then given by
\begin{align}
\label{eq:RIXSOnExpPolBasis}
\frac{\partial^2 \sigma}{\partial \Omega \partial \omega}  \propto  -\mathrm{Im} \brakettt{\Psi_0}{\left(R_{\omega_{\mathrm{in}},\mathbf{q}'}^{\boldsymbol{\hat{\epsilon}}_{\mathrm{in}}\boldsymbol{\hat{\epsilon}}_{\mathrm{out}}}\right)^{\dagger} G(\omega) R_{\omega_{\mathrm{in}},\mathbf{q}}^{\boldsymbol{\hat{\epsilon}}_{\mathrm{in}}\boldsymbol{\hat{\epsilon}}_{\mathrm{out}}}}{\Psi_0},
\end{align}
with $\mathbf{q}=\mathbf{q\,}'$ for momentum-conserving experiments.

It is useful to separate the influence of material properties from that of the experimental geometry in RIXS spectra. In (non-)linear optics, this is commonly achieved using susceptibility tensors, a formalism that can be directly applied to X-ray spectroscopy\cite{armstrong_interactions_1962,maker_study_1965,duboisset_nonlinear_2025}. RIXS, being equivalent to a resonant Raman process at X-ray frequencies, can be described by 4-point correlation function or RIXS tensor $\chi^{(3)}(\omega_{\mathrm{in}},\omega,\mathbf{q})$. In the dipole approximation, this tensor explicitly depends on the photon energy and transferred momentum but is independent of the light polarization, magnitude of the involved photon momenta and specific experimental setup. The polarization dependence of a given RIXS experiment enters through the contraction of the RIXS tensor with the light polarization:
\begin{align}
\label{eq:IntensityFromRIXSCarthesianRank4Tensor}
\frac{\partial^2 \sigma}{\partial \Omega \partial \omega}  &\propto  \\ \nonumber 
-\mathrm{Im} & \sum_{i,j,k,l} \boldsymbol{\hat{\epsilon}}_{\mathrm{in},i}^{\,*} \boldsymbol{\hat{\epsilon}}_{\mathrm{out},j} \, \chi^{(3)}_{i,j,k,l}(\omega_{\mathrm{in}},\omega,\mathbf{q}) \, \boldsymbol{\hat{\epsilon}}_{\mathrm{in},l} \, \boldsymbol{\hat{\epsilon}}_{\mathrm{out},k}^{\,*} ,
\end{align}
where the indices $i,j,k,l \in \{x,y,z\}$ correspond to the vector components of the polarization, and
\begin{align}
\label{eq:RIXSCarthesianRank4Tensor}
    \chi^{(3)}_{i,j,k,l}(\omega_{\mathrm{in}},\omega,\mathbf{q}) = \brakettt{\Psi_0}{\left(R_{\omega_{\mathrm{in}},\mathbf{q}}^{\boldsymbol{\hat{\epsilon}}_i\boldsymbol{\hat{\epsilon}}_j}\right)^{\dagger} G(\omega) R_{\omega_{\mathrm{in}},\mathbf{q}}^{\boldsymbol{\hat{\epsilon}}_l\boldsymbol{\hat{\epsilon}}_k}}{\Psi_0}.
\end{align}

The RIXS tensor in Eq.~\eqref{eq:RIXSCarthesianRank4Tensor} is a rank-4 Cartesian tensor in three dimensions, as it depends on four polarization components: two for the incoming light and two for the outgoing light. In Eq.~\eqref{eq:RIXSOnExpPolBasis}, the polarization vectors $\boldsymbol{\hat{\epsilon}}_{\mathrm{in}}$ and $\boldsymbol{\hat{\epsilon}}_{\mathrm{out}}$ appear naturally in the RIXS transition operators $R_{\omega_{\mathrm{in}},\mathbf{q}}^{\boldsymbol{\hat{\epsilon}}_{\mathrm{in}}\boldsymbol{\hat{\epsilon}}_{\mathrm{out}}}$ and their Hermitian conjugates, ensuring that the same polarization components are contracted. As a result, Eq.~\eqref{eq:RIXSOnExpPolBasis} might appear to involve only the diagonal elements of the RIXS tensor. However, when expanding the double differential cross section in a fixed basis, necessary to separate the polarization dependence from the RIXS tensor, the off-diagonal elements also become significant. Thus, in Eq.~\eqref{eq:RIXSCarthesianRank4Tensor}, it is essential to retain tensor components where $\boldsymbol{\hat{\epsilon}}_{i(j)}$ differs from $\boldsymbol{\hat{\epsilon}}_{l(k)}$.

\section{\label{sec:Symmetry}Symmetric coupling of tensor elements}

The RIXS tensor has $3^4=81$ complex components that describe the RIXS spectrum for all possible polarization states. In low-symmetry materials, all 81 components can be distinct and linearly independent. However, for crystals with symmetry, many of these components vanish, while others are related by symmetry constraints. To describe the symmetry properties of the RIXS tensor, it is useful to couple the polarization of the incoming and outgoing light. This results in a rank-2, 9-dimensional Cartesian tensor. A natural basis for this tensor is the direct Cartesian product of $\boldsymbol{\hat{\epsilon}}_{\mathrm{in}}$ and $\boldsymbol{\hat{\epsilon}}_{\mathrm{out}}^{\,*}$:
\begin{align}
    \hat{\mathbf{e}} &= \{\boldsymbol{\hat{\epsilon}}_{\mathrm{in}} \otimes \boldsymbol{\hat{\epsilon}}_{\mathrm{out}}^{\,*}\}\\ \nonumber
    &=\{xx^*,xy^*,xz^*,yx^*,yy^*,yz^*,zx^*,zy^*,zz^*\},
\end{align}
where the first index refers to the $x$, $y$, or $z$ component of $\boldsymbol{\hat{\epsilon}}_{\mathrm{in}}$, and the second index corresponds to the $x$, $y$, or $z$ component of $\boldsymbol{\hat{\epsilon}}^{\,*}_{\mathrm{out}}$.

With the coupled vector $\hat{\mathbf{e}} = \{\boldsymbol{\hat{\epsilon}}_{\mathrm{in}} \otimes \boldsymbol{\hat{\epsilon}}_{\mathrm{out}}^{\,*}\}$, the tensor can be expressed as a rank-2, 9-dimensional tensor. The scattered intensity is then given by
\begin{align}
\frac{\partial^2 \sigma}{\partial \Omega \partial \omega}  \propto -\mathrm{Im}  \sum_{a,b} \hat{\mathbf{e}}^{\,*}_{a}  \chi^{(3)}_{a,b}(\omega_{\mathrm{in}},\omega,\mathbf{q}) \, \hat{\mathbf{e}}_{b}.
\end{align}

The symmetry properties of point groups are determined by rotations and inversions. These properties are generally more intuitive when described using spherical tensors instead of Cartesian tensors. In Cartesian notation, a rank-1, 3-dimensional Cartesian tensor is equivalent to a spherical tensor of rank $l=1$. A spherical tensor of rank $l=1$ can be expressed in a basis of complex spherical harmonics $\{Y_{1,-1}(\theta,\phi),Y_{1,0}(\theta,\phi),Y_{1,1}(\theta,\phi)\}$. After a unitary basis transformation, it can also be expressed in a cubic harmonics basis $\{{x},{y},{z}\}$. The relation between these basis representations is given by
\begin{align} \nonumber
Y_{1,-1}(\theta,\phi) &= \frac{1}{2} \sqrt{\frac{3}{2 \pi}} e^{-\I\phi} \sin(\theta) &&= \frac{1}{2} \sqrt{\frac{3}{2 \pi}} (x - \I y), \\ \nonumber
Y_{1,\phantom{-}0}(\theta,\phi) &= \frac{1}{2} \sqrt{\frac{3}{\pi}} \cos(\theta) &&= \frac{1}{2} \sqrt{\frac{3}{\pi}} z, \\
Y_{1,\phantom{-}1}(\theta,\phi) &= -\frac{1}{2} \sqrt{\frac{3}{2 \pi}} e^{\I\phi} \sin(\theta) &&=-\frac{1}{2} \sqrt{\frac{3}{2 \pi}} (x + \I y),
\end{align}
and these relations can be used to transform the RIXS tensor into a basis of spherical tensors.

For Cartesian tensors, the dimension is always three in our discussion, while the rank determines the number of indices. In contrast, for spherical tensors, the rank corresponds to the angular momentum $l$ of the tensor representation, and the dimension is given by $2l+1$.

To couple $\boldsymbol{\hat{\epsilon}}_{\mathrm{in}}$ and $\boldsymbol{\hat{\epsilon}}^{*}_{\mathrm{out}}$, we can express the polarization vectors in a basis of complex spherical harmonics and apply Clebsch-Gordan coefficients for spherical vector coupling \cite{varshalovich_quantum_1988}:
\begin{align}
\{(\boldsymbol{\hat{\epsilon}}_{\mathrm{in}})^{(l_1)}_{m_1} \otimes (\boldsymbol{\hat{\epsilon}}^{*}_{\mathrm{out}})^{(l_2)}_{m_2} \}^{(l)}_{m} = \sum_{m_1,m_2}\nonumber& \brakett{l_1,m_1;l_2,m_2}{l,m}\\&\times(\boldsymbol{\hat{\epsilon}}_{\mathrm{in}})^{(l_1)}_{m_1}(\boldsymbol{\hat{\epsilon}}^{*}_{\mathrm{out}})^{(l_2)}_{m_2}.
\end{align}
For dipole interactions, we have $l_1=l_2=1$, leading to $0\leq l \leq 2$. This yields a total of $\sum_{l\in\{0,1,2\}} (2l+1) = 1+3+5=9$ basis states for the coupled polarization vectors $ \boldsymbol{\hat{\epsilon}}_{\mathrm{in}}$ and $ \boldsymbol{\hat{\epsilon}}_{\mathrm{out}}^{\,*}$. Explicitly, this basis is
\begin{align}
    \hat{\mathbf{e}} &= \{ \boldsymbol{\hat{\epsilon}}_{\mathrm{in}} \otimes \boldsymbol{\hat{\epsilon}}_{\mathrm{out}}^{\,*} \}\\ \nonumber
    &=\{ \hat{\mathbf{e}}^{(0)},\hat{\mathbf{e}}^{(1)}_{R_x},\hat{\mathbf{e}}^{(1)}_{R_y},\hat{\mathbf{e}}^{(1)}_{R_z}, \hat{\mathbf{e}}^{(2)}_{x^2-y^2}, \hat{\mathbf{e}}^{(2)}_{z^2}, \hat{\mathbf{e}}^{(2)}_{yz}, \hat{\mathbf{e}}^{(2)}_{xz}, \hat{\mathbf{e}}^{(2)}_{xy} \},
\end{align}
with
\begin{align} \nonumber
	\hat{\mathbf{e}}^{(0)} &= -\frac{\hat{\epsilon}_{\text{in},x} \hat{\epsilon}_{\text{out},x}^*}{\sqrt{3}} - \frac{\hat{\epsilon}_{\text{in},y} \hat{\epsilon}_{\text{out},y}^*}{\sqrt{3}} - \frac{\hat{\epsilon}_{\text{in},z} \hat{\epsilon}_{\text{out},z}^*}{\sqrt{3}}, 
    \\ \nonumber
	\hat{\mathbf{e}}^{(1)}_{R_x}&= \frac{i \hat{\epsilon}_{\text{in},y} \hat{\epsilon}_{\text{out},z}^*}{\sqrt{2}} - \frac{i \hat{\epsilon}_{\text{in},z} \hat{\epsilon}_{\text{out},y}^*}{\sqrt{2}}, \\\nonumber
	\hat{\mathbf{e}}^{(1)}_{R_y} &= \frac{i \hat{\epsilon}_{\text{in},z} \hat{\epsilon}_{\text{out},x}^*}{\sqrt{2}} - \frac{i \hat{\epsilon}_{\text{in},x} \hat{\epsilon}_{\text{out},z}^*}{\sqrt{2}}, \\\nonumber
	\hat{\mathbf{e}}^{(1)}_{R_z} &= \frac{i \hat{\epsilon}_{\text{in},x} \hat{\epsilon}_{\text{out},y}^*}{\sqrt{2}} - \frac{i \hat{\epsilon}_{\text{in},y} \hat{\epsilon}_{\text{out},x}^*}{\sqrt{2}},
    \\ \nonumber
	\hat{\mathbf{e}}^{(2)}_{x^2-y^2} &= \frac{\hat{\epsilon}_{\text{in},x} \hat{\epsilon}_{\text{out},x}^*}{\sqrt{2}} - \frac{\hat{\epsilon}_{\text{in},y} \hat{\epsilon}_{\text{out},y}^*}{\sqrt{2}},\\\nonumber
	\hat{\mathbf{e}}^{(2)}_{z^2} &= -\frac{\hat{\epsilon}_{\text{in},x} \hat{\epsilon}_{\text{out},x}^*}{\sqrt{6}} - \frac{\hat{\epsilon}_{\text{in},y} \hat{\epsilon}_{\text{out},y}^*}{\sqrt{6}} + \sqrt{\frac{2}{3}} \hat{\epsilon}_{\text{in},z} \hat{\epsilon}_{\text{out},z}^*, \\\nonumber
	\hat{\mathbf{e}}^{(2)}_{yz} &= \frac{\hat{\epsilon}_{\text{in},y} \hat{\epsilon}_{\text{out},z}^*}{\sqrt{2}} + \frac{\hat{\epsilon}_{\text{in},z} \hat{\epsilon}_{\text{out},y}^*}{\sqrt{2}}, \\\nonumber
	\hat{\mathbf{e}}^{(2)}_{xz} &= \frac{\hat{\epsilon}_{\text{in},x} \hat{\epsilon}_{\text{out},z}^*}{\sqrt{2}} + \frac{\hat{\epsilon}_{\text{in},z} \hat{\epsilon}_{\text{out},x}^*}{\sqrt{2}}, \\
	\hat{\mathbf{e}}^{(2)}_{xy} &= \frac{\hat{\epsilon}_{\text{in},y} \hat{\epsilon}_{\text{out},x}^*}{\sqrt{2}} + \frac{\hat{\epsilon}_{\text{in},x} \hat{\epsilon}_{\text{out},y}^*}{\sqrt{2}}. \label{eq:sphericalkubiktensorbasis}
\end{align}

Here, we use cubic harmonics as a basis for our spherical tensors, which is particularly useful for cubic, tetragonal, or orthorhombic point groups.

The RIXS tensor is now represented as a $9\times9$ matrix with complex entries. Depending on symmetry, many of these elements are related or vanish. In spherical symmetry, the tensor is diagonal with three independent elements corresponding to different angular momentum components. When the local point group is reduced, one can use point-group branching ratios to determine which irreducible representations contribute to the RIXS tensor. If multiple diagonal elements belong to the same irreducible representation, non counting tensor elements belonging to the same multidimensional irreducible representation, then off-diagonal elements between them also become nonzero. 


\section{\label{sec:NumExample1}Numerical examples in some non-magnetic point groups}

\begin{figure*}
    \centering
    \includegraphics{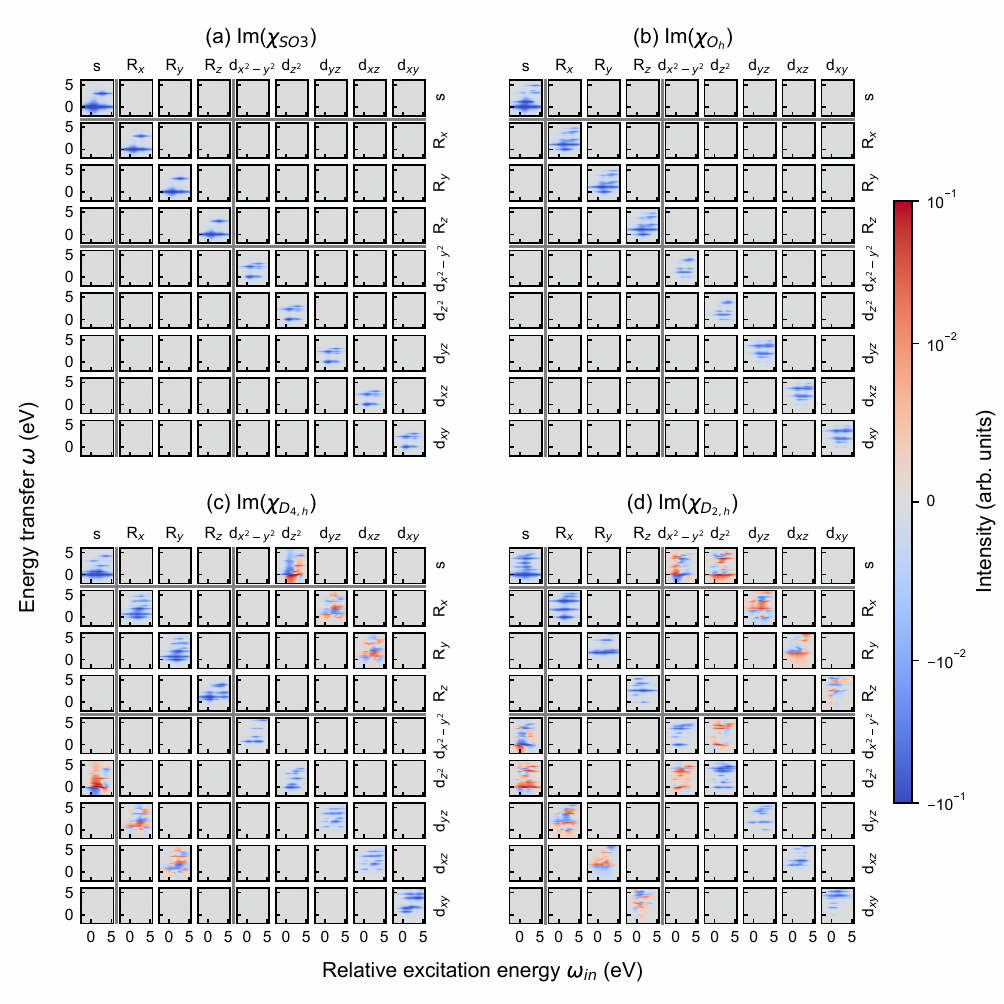}
    \caption{\label{fig:ChiNiONonmagnetic} Imaginary part of the 2p$_{3/2}$3d RIXS tensor for systems with (a)~SO(3), (b)~$O_h$, (c)~$D_{4h}$, and (d)~$D_{2h}$ point group symmetry. Each tensor consists of 81 energy transfer over excitation energy intensity maps. Excitations energies are expressed relatively to the binding energy of Ni 2p$_{1/2}$ states. A symmetric logarithmic color scale is used to plot the values, with a linear regime between -0.01 and 0.01. The tensors are represented in a cubic harmonic basis, as defined in Eq.~(\ref{eq:sphericalkubiktensorbasis}). The dark grey lines separate blocks with different angular momentum l.} 
\end{figure*}

Figure~\ref{fig:ChiNiONonmagnetic} shows the imaginary part of the 2p$_{3/2}$3d RIXS tensor for a Ni$^{2+}$ ion in a crystal field with spherical ($SO(3)$), cubic ($O_h$), tetragonal ($D_{4h}$), or orthorhombic ($D_{2h}$) point group symmetry. The calculations were performed using \textsc{Quanty} \cite{QuantyWebpage,haverkort_quanty_2016,ackermann_quanty_2024}, and the input scripts are available in the online supplementary materials. The model includes atomic Coulomb interactions, spin-orbit coupling, and the crystal field, with reasonable parameters. We present $\mathrm{Im} (\chi_{a,b}^{(3)}(\omega_{\mathrm{in}},\omega))$ for $a,b \in \{ \hat{\mathbf{e}}^{(0)},\hat{\mathbf{e}}^{(1)}_{R_x},\hat{\mathbf{e}}^{(1)}_{R_y},\hat{\mathbf{e}}^{(1)}_{R_z}, \hat{\mathbf{e}}^{(2)}_{x^2-y^2}, \hat{\mathbf{e}}^{(2)}_{z^2}, \hat{\mathbf{e}}^{(2)}_{yz}, \hat{\mathbf{e}}^{(2)}_{xz}, \hat{\mathbf{e}}^{(2)}_{xy} \}$, yielding a matrix of 81 RIXS spectra for each point group. Each element of $\chi^{(3)}(\omega_{\mathrm{in}},\omega)$ consists of a 2-dimensional map as a function of the excitation energy $\omega_{\mathrm{in}}$ and energy transfer $\omega$. Dependency on $\mathbf{q}$ is absent, since only a single ion is considered. We scan excitation energies from 2.5~eV below the $L_3$ binding energy to 5.5~eV above it, and energy transfer $\omega$ from -2~eV to 6~eV.

Figure~\ref{fig:ChiNiONonmagnetic}(a) shows $\chi^{(3)}(\omega_{\mathrm{in}},\omega)$ for a Ni$^{2+}$ ion in spherical symmetry. In this case, the tensor is diagonal and all elements belonging to the same angular momentum representation are equivalent. As a result, only three distinct spectra define the entire RIXS tensor in spherical symmetry. 

\begin{table}[tbh]
\begin{ruledtabular}
\centering
\begin{tabular}{l||c|ccc|ccccc}
		& s & $R_x$ & $R_y$ & $R_z$ & $d_{x^2 - y^2}$ & $d_{z^2}$ & $d_{yz}$ & $d_{xz}$ & $d_{xy}$ \\
		\hline   
		SO(3) \rule{0pt}{10pt}     & s & \multicolumn{3}{c|}{Rot}  & \multicolumn{5}{c}{d}  \\ 
		O$_h$         & $a_{1g}$ & \multicolumn{3}{c|}{$t_{1g}$}  & \multicolumn{2}{c|}{$e_{g}$} & \multicolumn{3}{c}{$t_{2g}$} \\ 
		D$_{4h}$   & $a_{1g}$ & \multicolumn{2}{c|}{$e_{g}$} & $a_{2g}$        & \multicolumn{1}{c|}{$b_{1g}$}    & \multicolumn{1}{c|}{$a_{1g}$} & \multicolumn{2}{c|}{$e_{g}$} & $b_{2g}$ \\ 
		D$_{2h}$   & \multicolumn{1}{c|}{$a_{g}$}  & \multicolumn{1}{c|}{$b_{3g}$} & \multicolumn{1}{c|}{$b_{2g}$} &$b_{1g}$ & \multicolumn{1}{c|}{$a_g$} & \multicolumn{1}{c|}{$a_g$} & \multicolumn{1}{c|}{$b_{3g}$} & \multicolumn{1}{c|}{$b_{2g}$} & $b_{1g}$
	\end{tabular}
\end{ruledtabular}
\caption{\label{tab:branching}Branching rules for the irreducible representations of the coupled angular momenta $l \in\{0,1,2\}$ for the point groups considered in Figure~\ref{fig:ChiNiONonmagnetic}.}
\end{table}

To understand how the RIXS tensor changes for different point groups, we use the branching rules for irreducible representations listed in Table~\ref{tab:branching}. When transitioning from a spherical atom to an atom in a cubic environment, the $l=0$ representation becomes an $a_{1g}$ representation, the $l=1$ axial vector $R$ becomes a $t_{1g}$ representation and the $l=2$ representation branches into $t_{2g}$ and $e_g$ representations. Figure~\ref{fig:ChiNiONonmagnetic}(b) shows the RIXS tensor for a Ni$^{2+}$ ion in a cubic crystal field of 1.1~eV, the experimental value for NiO. In the cubic harmonic basis, the tensor is diagonal and consists of four distinct entries corresponding to the irreducible representations of $l=0$, $l=1$, and $l=2$.

Reducing the symmetry from cubic to tetragonal causes further branching of the representations. The cubic $a_{1g}$ representation remains $a_{1g}$, the $t_{1g}$ representation splits into $e_g$ and $a_{2g}$, the $e_g$ representation branches into $b_{1g}$ and $a_{1g}$, and the $t_{2g}$ representation splits into $e_g$ and $b_{2g}$, as summarized in Table~\ref{tab:branching}. Figure~\ref{fig:ChiNiONonmagnetic}(c) shows that, as expected, the diagonal elements corresponding to different irreducible representations diverge. Additionally, off-diagonal elements appear. Whenever the same representation appears more than once on the diagonal, the off-diagonal elements between the corresponding entries can also be nonzero.

Furthermore, we observe that the off-diagonal elements are neither symmetric nor antisymmetric. Even in crystals with time-reversal symmetry, the RIXS experimental setup does not preserve time-reversal symmetry. This can be demonstrated with a simple example. Consider a tetragonal crystal with a Cu$^{2+}$ ion in a $d^9$ configuration, where a hole resides in the $3d_{x^2-y^2}$ orbital. An X-ray photon with $x$-polarized light can excite an electron from the $2p_x$ state to the $3d_{x^2-y^2}$ orbital. The intermediate $2p^5 3d^{10}$ state can then decay by emitting a $z$-polarized photon, creating a hole in the $3d_{xz}$ orbital. Thus, the system allows scattering from an $x$-polarized photon to a $z$-polarized photon. However, the inverse process—scattering from $z$ to $x$ polarization—has zero cross-section because a $z$-polarized photon cannot excite an electron from the $2p$ shell to the $3d_{x^2-y^2}$ orbital. Inverting the scattering process is only possible when starting from the excited state, but the RIXS tensor describes scattering from the ground state, meaning that scattering from $x$ to $z$ is not related to scattering from $z$ to $x$ by symmetry arguments.
Further reducing the symmetry to the $D_{2h}$ point-group results in the RIXS tensor shown in Fig.~\ref{fig:ChiNiONonmagnetic}(d). The irreducible representations for this case are listed in Table~\ref{tab:branching}. Since all cubic harmonics belong to 1-dimensional irreducible representations, all nine diagonal elements of $\chi^{(3)}(\omega_{\mathrm{in}},\omega)$ are distinct. Additionally, three elements belong to the $a_{1g}$ representation, and two each belong to the $b_{1g}$, $b_{2g}$, and $b_{3g}$ representations. This introduces another $2 \times 3 + 2 \times 3 = 12$ nonzero matrix elements. In total, there are 21 linearly independent RIXS spectra that can be measured as a function of polarization in an orthorhombic system.

\section{\label{sec:NumExample2}Numerical examples in some magnetic point groups}

\begin{figure*}
    \centering
    \includegraphics{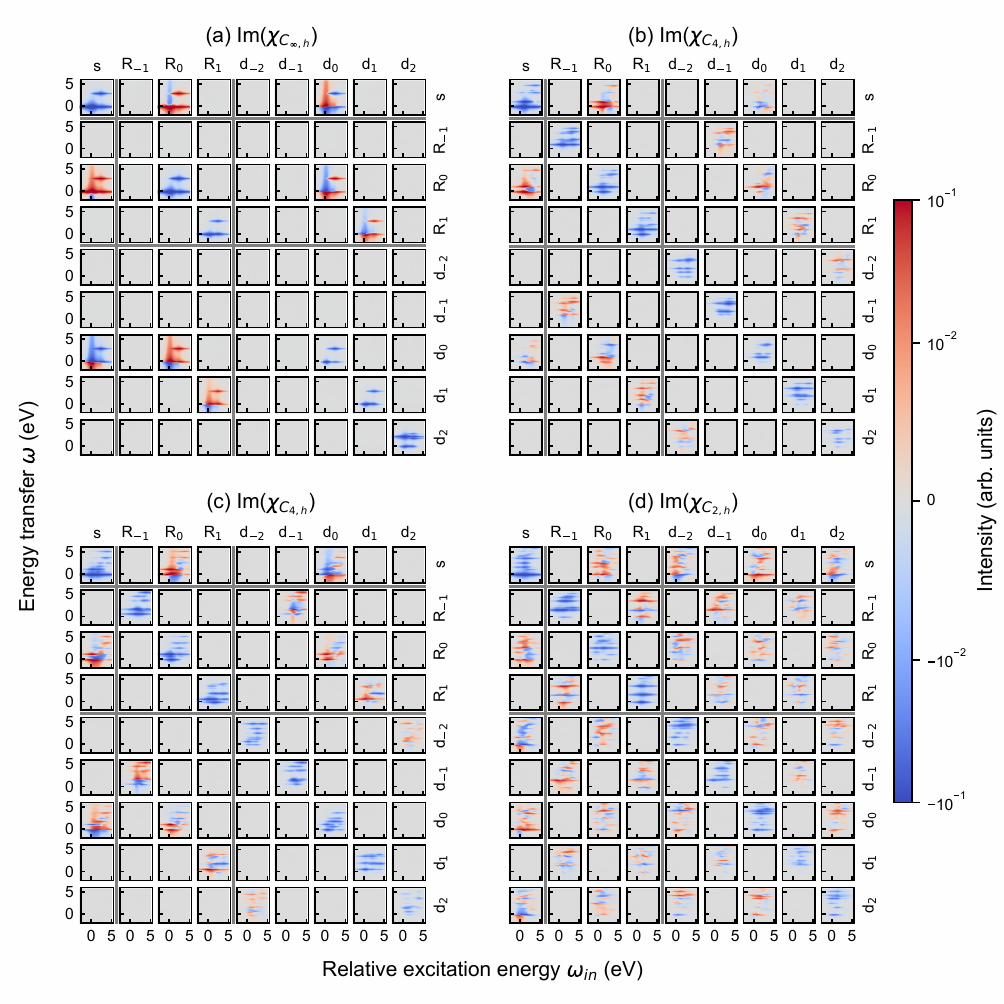}
    \caption{\label{fig:FigTensorLowerPGMagneticKubik} Imaginary part of the 2p$_{3/2}$3d RIXS tensor for systems with (a)~SO(3), (b)~$O_h$, (c)~$D_{4h}$, and (d)~$D_{2h}$ point group symmetry in the presence of a small but finite external magnetic field $B_{\mathrm{ext}} || \mathbf{z}$. Only the resulting single degenerate ground state is considered. Each tensor consists of 81 energy transfer over excitation energy intensity maps. Excitations energies are expressed relatively to the binding energy of Ni 2p$_{1/2}$ states. A symmetric logarithmic color scale is used to plot the values, with a linear regime between -0.01 and 0.01. The tensors are represented in a spherical harmonic basis, as defined in Eq.~(\ref{eq:sphericalkubiktensorbasis}). Table~\ref{tab:branchingWithBext} lists the branching rules for the unitary magnetic subgroups involved.  The dark grey lines separate blocks with different angular momentum l.}
\end{figure*}

Figure~\ref{fig:FigTensorLowerPGMagneticKubik} shows numerical results for the RIXS tensor for spherical (a), cubic (b), tetragonal (c), and orthorhombic (d) point group symmetries in the presence of an additional external magnetic field along the $z$ direction. To determine which tensor elements remain nonzero, we analyze the branching of the RIXS tensor representations under symmetry reduction. 

The $SO(3)$ point group reduces to $\infty/mm'$ when a magnetic field is applied along the $z$ axis. This group consists of an infinite-order rotation axis $C_{\infty}$ about the $z$ direction, a horizontal mirror plane ($m$), and a vertical mirror plane combined with time-reversal symmetry ($m'$). As we are interested in the RIXS tensor for a magnetized sample, time-reversal symmetry relates the RIXS tensor for a sample magnetized along the $+z$ direction to that of one magnetized along $-z$. However, for a specific magnetization direction, we can consider the unitary subgroup that excludes time-reversal symmetry. Starting from the non-magnetic point groups $SO(3)$, $O_h$, $D_{4h}$, and $D_{2h}$, we obtain the following magnetic groups and their unitary subgroups when a magnetization is introduced along the $z$ direction:
\begin{alignat}{5}
\text{SO(3)} & \, \rightarrow\, & \infty/mm' & \, \triangleright \, & \text{C}_{\infty h}, \nonumber\\
\text{O}_{h} & \, \rightarrow\, & 4/mm'm' & \, \triangleright \, & \text{C}_{4h}, \nonumber\\
\text{D}_{4h} & \, \rightarrow\, & 4/mm'm' & \, \triangleright \, & \text{C}_{4h}, \nonumber\\
\text{D}_{2h} & \, \rightarrow\, & 2/mm'm' & \, \triangleright \, & \text{C}_{2h}.
\label{eq:PGwithBext}
\end{alignat}

We now examine the irreducible representations within these unitary subgroups. Care must be taken in choosing the fully irreducible representations, as certain point groups, such as $C_{\infty h}$ and $C_{4h}$, contain complex irreducible representations. In many character tables, two complex conjugate representations are combined into a reducible $e_g$ representation. The RIXS tensor on this reducible basis is however not diagonal. In our case, the pairs $(R_x, R_y)$, $(d_{x^2-y^2}, d_{xy})$, and $(d_{xz}, d_{yz})$ each form such $e_g$ representations. By transforming the spherical tensor-coupled polarization basis from cubic harmonics to complex spherical harmonics, defined as
\begin{align}
R_{\pm1} &= \sqrt{\frac{1}{2}} (\mp R_x - \I R_y), \nonumber \\
d_{\pm1} &= \sqrt{\frac{1}{2}} (\mp d_{xz} - \I d_{yz}), \nonumber \\
d_{\pm2} &= \sqrt{\frac{1}{2}} (d_{x^2-y^2} \pm \I d_{xy}).
\end{align}
We return to a basis of irreducible representations on which the RIXS tensor is diagonal within that subspace.

\begin{table}
\begin{ruledtabular}
\centering
\begin{tabular}{l||c|ccc|ccccc}
    & s & $R_{-1}$ & $R_{0}$ & $R_{1}$ & $d_{-2}$ & $d_{-1}$ & $d_{0}$ & $d_{1}$ & $d_{2}$ \\
    \hline     
    C$_{\infty,h}\rule{0pt}{10pt}$ &$\Sigma_g$ &  $\Pi_g^-$ & $\Sigma_g$ & $\Pi_g^+$ & $\Delta_g^-$ & $\Pi_g^-$ & $\Sigma_g$ & $\Pi_g^+$ & $\Delta_g^+$ \\ 
    C$_{4h}$            & $a_{g}$ & $e_{g}^-$ & $a_{g}$ & $e_{g}^+$       & $b_g$        & $e_{g}^-$ & $a_g$ & $e_{g}^+$ & $b_{g}$ \\ 
    C$_{2h}$            & $a_{g}$ & $b_{g}$ & $a_{g}$ & $b_{g}$        & $a_g$        & $b_{g}$ & $a_g$ & $b_{g}$ & $a_{g}$ \\ 
\end{tabular}
\end{ruledtabular}
\caption{\label{tab:branchingWithBext}Irreducible representations of the relevant magnetic unitary subgroups obtained by introducing a magnetic field in the $z$ direction to the cubic, tetragonal, and orthorhombic groups listed in Table~\ref{tab:branching}.}
\end{table}

Table~\ref{tab:branchingWithBext} lists the irreducible representations for the basis functions used in the RIXS tensor $\chi^{(3)}(\omega_{\mathrm{in}},\omega,\mathbf{q})$ in the presence of a magnetic field. All representations are 1-dimensional, meaning that all nine diagonal elements of the tensor can be distinct. This is indeed observed in Fig.~\ref{fig:FigTensorLowerPGMagneticKubik}(b), (c), and (d), where we start from an $O_h$, $D_{4h}$, or $D_{2h}$ crystal point group and introduce a magnetic field along the $z$ direction.

In the case of spherical symmetry (Fig.~\ref{fig:FigTensorLowerPGMagneticKubik}(a)), several diagonal elements vanish. This occurs due to the special case of a Hund's-rule high-spin ground state, where $L$, $S$, and $J$ are maximal, and $J_z = -J$ is minimal. Additionally, in spherical symmetry, the total angular momentum of the electronic system is conserved. As a result, transitions that reduce the total angular momentum along $z$ cannot occur, meaning that all representations with $m < 0$ vanish. This is a unique property of spherical symmetry, as in all crystalline environments, angular momentum is not conserved, since electrons can scatter off the crystal field, transferring momentum to the lattice.

For an orthorhombic crystal with a magnetic field along one of its symmetry axes, tensor elements with odd $m$ belong to the $b_g$ representation, while those with even $m$ belong to the $a_g$ representation. Consequently, the RIXS tensor retains  41 nonzero elements out of 81 possible elements, all of which are linearly independent.

\section{\label{sec:AvvEpsOut}Averaging over the outgoing polarization}

In most experiments, the polarization of the scattered light is not analyzed. In such cases, one must average over all possible polarization states of the scattered photon, constrained to be perpendicular to the wave vector $\mathbf{k}_{\mathrm{out}}$. Using Eq.~\eqref{eq:poldefinition}, we express the outgoing polarization vector in terms of the basis vectors $\boldsymbol{\hat{\sigma}}_{\mathrm{out}}$ and $\boldsymbol{\hat{\pi}}_{\mathrm{out}}$, which are determined in the crystal coordinate system by Eq.~\eqref{eq:polarizationdef}. The outgoing polarization is written as $\boldsymbol{\hat{\epsilon}}_{\mathrm{out}} = \boldsymbol{\hat{\epsilon}}_{\mathrm{out}}(\alpha_{\mathrm{out}},\beta_{\mathrm{out}})$, where $\alpha_{\mathrm{out}}$ controls the ratio between $\sigma$- and $\pi$-polarized light, and $\beta_{\mathrm{out}}$ defines the phase difference between the two polarization components.

Without using a tensor formulation, one would need to compute the RIXS spectrum for each pair of values $\alpha_{\mathrm{out}}$ and $\beta_{\mathrm{out}}$ and subsequently average over all spectra. Instead of performing multiple RIXS calculations, one can compute the RIXS tensor once and average over specific tensor elements. Using Eq.~\eqref{eq:sphericalkubiktensorbasis}, we couple the incoming and outgoing polarization vectors to define $\hat{\bf{e}}(\alpha_{\mathrm{out}},\beta_{\mathrm{out}}) = \{\boldsymbol{\hat{\epsilon}}_{\mathrm{in}}^* \otimes \boldsymbol{\hat{\epsilon}}_{\mathrm{out}}(\alpha_{\mathrm{out}},\beta_{\mathrm{out}})\}$. The RIXS intensity without a polarization analyzer for the outgoing light is then given by
\begin{widetext}
\begin{align}
\label{eq:nopolfilterout}
\frac{\partial^2 \sigma}{\partial \Omega \partial \omega}  \propto  -\mathrm{Im}\bigg( &\frac{1}{4 \pi^2} \int_0^{2\pi} \int_0^{2\pi} \sum_{a,b} \big(\hat{\bf{e}}_{\epsilon_{\mathrm{in}},\epsilon_{\mathrm{out}}(\alpha_{\mathrm{out}},\beta_{\mathrm{out}})}^{*}\big)_{a}  \chi^{(3)}_{a,b}(\omega_{\mathrm{in}},\omega,\mathbf{q}) \, \big(\hat{\bf{e}}_{\epsilon_{\mathrm{in}},\epsilon_{\mathrm{out}}(\alpha_{\mathrm{out}},\beta_{\mathrm{out}})}\big)_{b} \, \mathrm{d} \alpha_{\mathrm{out}} \, \mathrm{d} \beta_{\mathrm{out}}\bigg) \\ \nonumber
=-\mathrm{Im}\bigg(&\frac{1}{4 \pi^2} \int_0^{2\pi} \int_0^{2\pi} \cos^2(\alpha_{\mathrm{out}})^2 \, \mathrm{d} \alpha_{\mathrm{out}} \, \mathrm{d} \beta_{\mathrm{out}}\sum_{a,b}\big(\hat{\bf{e}}_{\epsilon_{\mathrm{in}},\boldsymbol{\hat{\pi}}_{\mathrm{out}}}^{*}\big)_{a}  \chi^{(3)}_{a,b}(\omega_{\mathrm{in}},\omega,\mathbf{q}) \, \big(\hat{\bf{e}}_{\epsilon_{\mathrm{in}},\boldsymbol{\hat{\pi}}_{\mathrm{out}}}\big)_{b} \, \mathrm{d} \alpha_{\mathrm{out}} \, \mathrm{d} \beta_{\mathrm{out}}\\\nonumber
+&\frac{1}{4 \pi^2} \int_0^{2\pi} \int_0^{2\pi} \sin^2(\alpha_{\mathrm{out}}) \, \mathrm{d} \alpha_{\mathrm{out}} \, \mathrm{d} \beta_{\mathrm{out}}\sum_{a,b}\big(\hat{\bf{e}}^{*}_{\epsilon_{\mathrm{in}},\boldsymbol{\hat{\sigma}}_{\mathrm{out}}}\big)_{a}  \chi^{(3)}_{a,b}(\omega_{\mathrm{in}},\omega,\mathbf{q}) \, \big(\hat{\bf{e}}_{\epsilon_{\mathrm{in}},\boldsymbol{\hat{\sigma}}_{\mathrm{out}}}\big)_{b}\\\nonumber
+&\frac{1}{4 \pi^2} \int_0^{2\pi} \int_0^{2\pi} \cos(\alpha_{\mathrm{out}})\sin(\alpha_{\mathrm{out}})e^{-i\beta_{\mathrm{out}}} \, \mathrm{d} \alpha_{\mathrm{out}} \, \mathrm{d} \beta_{\mathrm{out}}\sum_{a,b}\big(\hat{\bf{e}}^{*}_{\epsilon_{\mathrm{in}},\boldsymbol{\hat{\pi}}_{\mathrm{out}}}\big)_{a}  \chi^{(3)}_{a,b}(\omega_{\mathrm{in}},\omega,\mathbf{q}) \, \big(\hat{\bf{e}}_{\epsilon_{\mathrm{in}},\boldsymbol{\hat{\sigma}}_{\mathrm{out}}}\big)_{b}\\\nonumber
+&\frac{1}{4 \pi^2} \int_0^{2\pi} \int_0^{2\pi} \cos(\alpha_{\mathrm{out}})\sin(\alpha_{\mathrm{out}})e^{i\beta_{\mathrm{out}}} \, \mathrm{d} \alpha_{\mathrm{out}} \, \mathrm{d} \beta_{\mathrm{out}}\sum_{a,b}\big(\hat{\bf{e}}^{*}_{\epsilon_{\mathrm{in}},\boldsymbol{\hat{\sigma}}_{\mathrm{out}}}\big)_{a}  \chi^{(3)}_{a,b}(\omega_{\mathrm{in}},\omega,\mathbf{q}) \, \big(\hat{\bf{e}}_{\epsilon_{\mathrm{in}},\boldsymbol{\hat{\pi}}_{\mathrm{out}}}\big)_{b}\bigg)\\ \nonumber
=-\mathrm{Im}\bigg(&\frac{1}{2} \sum_{a,b}\big(\hat{\bf{e}}^{*}_{\epsilon_{\mathrm{in}},\boldsymbol{\hat{\pi}}_{\mathrm{out}}}\big)_{a}  \chi^{(3)}_{a,b}(\omega_{\mathrm{in}},\omega,\mathbf{q}) \, \big(\hat{\bf{e}}_{\epsilon_{\mathrm{in}},\boldsymbol{\hat{\pi}}_{\mathrm{out}}}\big)_{b}+\frac{1}{2} \sum_{a,b}\big(\hat{\bf{e}}^{*}_{\epsilon_{\mathrm{in}},\boldsymbol{\hat{\sigma}}_{\mathrm{out}}}\big)_{a}  \chi^{(3)}_{a,b}(\omega_{\mathrm{in}},\omega,\mathbf{q}) \, \big(\hat{\bf{e}}_{\epsilon_{\mathrm{in}},\boldsymbol{\hat{\sigma}}_{\mathrm{out}}}\big)_{b} \bigg).
\end{align}
\end{widetext}
The observed spectrum is thus the sum between the process with $\boldsymbol{\hat{\pi}}_{out}$ and the one with $\boldsymbol{\hat{\sigma}}_{out}$, while cross terms are averaged out by integration. This result is in agreement with experimental observations \cite{fumagalli_polarization-resolved_2019}.

To illustrate how Eq.~\eqref{eq:nopolfilterout} is applied, we consider a non-magnetic cubic crystal with $O_h$ point group symmetry under different experimental geometries. We examine the scattering of either $\sigma$- or $\pi$-polarized incident light for a scattering angle of $2\theta = 90^\circ$. We consider two experimental configurations: 
1. $\mathbf{k}_{\mathrm{in}}=\{1,0,0\}$ and $\mathbf{k}_{\mathrm{out}}=\{0,1,0\}$.
2. $\mathbf{k}_{\mathrm{in}}=\{\sqrt{1/2},\sqrt{1/2},0\}$ and $\mathbf{k}_{\mathrm{out}}=\{-\sqrt{1/2},\sqrt{1/2},0\}$.

The first case aligns the momentum vectors with the $C_4$ symmetry axis of the crystal, while the second aligns them with the $C_{2d}$ symmetry axis. We assume no polarization filter for the outgoing light. The relevant polarization vectors are:
\begin{align}  \nonumber
\boldsymbol{\hat{\sigma}}^{C_4}_{\mathrm{in/out}} &= \{0,0,1\}, \\ \nonumber
\boldsymbol{\hat{\pi}}^{C_4}_{\mathrm{in}} &= \{0,-1,0\}, \\ \nonumber
\boldsymbol{\hat{\pi}}^{C_4}_{\mathrm{out}} &= \{1,0,0\}, \\ \nonumber
\boldsymbol{\hat{\sigma}}^{C_{2d}}_{\mathrm{in/out}} &= \{0,0,1\}, \\ \nonumber
\boldsymbol{\hat{\pi}}^{C_{2d}}_{\mathrm{in}} &= \{\sqrt{1/2},-\sqrt{1/2},0\}, \\
\boldsymbol{\hat{\pi}}^{C_{2d}}_{\mathrm{out}} &= \{\sqrt{1/2},\sqrt{1/2},0\}.
\label{eq:polexample}
\end{align}
To compute the RIXS intensity, we consider the RIXS tensor $\chi$ and the coupled light polarization vector $\hat{\bf{e}} = \boldsymbol{\hat{\epsilon}}_{\mathrm{in}}^* \otimes \boldsymbol{\hat{\epsilon}}^{*}_{\mathrm{out}}$ and evaluate the scalar product $\hat{\bf{e}}^* \cdot \chi \cdot \hat{\bf{e}}$.

For a non-magnetic system with local cubic ($O_h$) symmetry, the nonzero elements of the RIXS tensor include $\chi_{a_{1g}}$, $\chi_{t_{1g}}$, $\chi_{e_{g}}$, and $\chi_{t_{2g}}$, as listed in Table~\ref{tab:branching}. In this symmetry, the nonzero elements of $\chi$ lie on the diagonal of the tensor, as exemplified in Fig.~\ref{fig:ChiNiONonmagnetic}(b).
For a given experiment, we express $\hat{\bf{e}}$ in crystal coordinates. Inserting these expressions into Eq.~\eqref{eq:sphericalkubiktensorbasis},we obtain:
\begin{align} \nonumber
    \hat{\bf{e}}_{\pi_{\textrm{in}},\pi_{\textrm{out}}}^{C_4} & = \bigg\{0,0, 0,\,\I \sqrt{\frac{1}{2}}\,,0, 0, \,0, 0,-\sqrt{\frac{1}{2}} \bigg\},\nonumber\\
    \hat{\bf{e}}_{\pi_{\textrm{in}},\sigma_{\textrm{out}}}^{C_4} & = \bigg\{0,-\I \sqrt{\frac{1}{2}}\,, 0, 0, 0,0,-\sqrt{\frac{1}{2}}\,,0, 0\bigg\},\nonumber\\
    \hat{\bf{e}}_{\sigma_{\textrm{in}},\pi_{\textrm{out}}}^{C_4} & = \bigg\{0, \, 0, \,\I \sqrt{\frac{1}{2}}\,,0, \, 0, 0,0,\sqrt{\frac{1}{2}}\,,0 \bigg\},\nonumber\\
    \hat{\bf{e}}_{\sigma_{\textrm{in}},\sigma_{\textrm{out}}}^{C_4} & = \bigg\{-\sqrt{\frac{1}{3}}\,,0, 0, 0, 0, \sqrt{\frac{2}{3}}\,,0,0, 0\bigg\}, \nonumber
    \\
    \hat{\bf{e}}_{\pi_{\textrm{in}},\pi_{\textrm{out}}}^{C_{2d}} & = \bigg\{0, \, 0, 0,\,\I \sqrt{\frac{1}{2}},\sqrt{\frac{1}{2}}\,, \, 0, 0, 0, 0\bigg\},\nonumber\\
    \hat{\bf{e}}_{\pi_{\textrm{in}},\sigma_{\textrm{out}}}^{C_{2d}} & = \bigg\{0,-\frac{\I}{2}, -\frac{\I}{2}, 0, 0,  \, 0, -\frac{1}{2},\frac{1}{2}, 0\bigg\},\nonumber\\
    \hat{\bf{e}}_{\sigma_{\textrm{in}},\pi_{\textrm{out}}}^{C_{2d}} & = \bigg\{0,-\frac{\I}{2}, \frac{\I}{2}, 0, 0,  \, 0, \frac{1}{2},\frac{1}{2}, 0\bigg\},\nonumber\\
    \hat{\bf{e}}_{\sigma_{\textrm{in}},\sigma_{\textrm{out}}}^{C_{2d}} & = \bigg\{-\sqrt{\frac{1}{3}}\,,0, 0, 0, 0,\sqrt{\frac{2}{3}}\,,0,0, 0\bigg\}.
    \\\nonumber
    \end{align}
We now use Eq.~(\ref{eq:nopolfilterout}) to compute the RIXS tensor for the given experimental configurations:
\begin{align} \nonumber
    \chi^{C_4}_{\boldsymbol{\hat{\pi}}_{\mathrm{in}}} &= \frac{1}{2} \chi_{t_{1g}} + \frac{1}{2} \chi_{{t_{2g}}}, \\ \nonumber
    \chi^{C_4}_{\boldsymbol{\hat{\sigma}}_{\mathrm{in}}} &= \frac{2}{12}\chi_{a_{1g}} +\frac{3}{12}\chi_{t_{1g}} + \frac{4}{12}\chi_{e_g} + \frac{3}{12}\chi_{t_{2g}}, \\ 
    \nonumber
    \chi^{C_{2d}}_{\boldsymbol{\hat{\pi}}_{\mathrm{in}}} &= \frac{1}{2} \chi_{t_{1g}}+\frac{1}{4} \chi_{e_g}  + \frac{1}{4} \chi_{{t_{2g}}}, \\ \nonumber \text{and} \\ \nonumber
    \chi^{C_{2d}}_{\boldsymbol{\hat{\sigma}}_{\mathrm{in}}} &= \frac{2}{12}\chi_{a_{1g}} + \frac{3}{12}\chi_{t_{1g}} + \frac{4}{12}\chi_{e_g} + \frac{3}{12}\chi_{t_{2g}}.
    \nonumber
\end{align}
For a given experimental geometry, one typically does not measure a single symmetry component of $\chi$, but rather a linear combination of multiple components. By performing measurements with different polarization settings, one can reconstruct the full tensor.

\section{\label{sec:AvvPowder}Powder averaging}

Many RIXS experiments, particularly those involving core-to-core transitions, are performed on powder samples. To compute the scattered intensity in such cases, one must average over all possible sample orientations.  This can be achieved by defining the spherically averaged RIXS tensor:
\begin{align}
\overline{\chi} = \frac{1}{8\pi^2} \int_0^{2 \pi} \!\!\!\!\! \int_0^{\pi} \!\!\!\! \int_0^{2 \pi} \!\!\!\! D(\phi,\theta,\psi)^{\dag} \chi D(\phi,\theta,\psi) \sin(\theta) \, \mathrm{d} \phi \, \mathrm{d} \theta \, \mathrm{d} \psi,
\label{eq:ChiIntegralAvg}
\end{align}
where $D$ is a rotation matrix acting on the RIXS tensor, and $\phi$, $\theta$, and $\psi$ are the Euler angles defining a general rotation. Transferred momentum dependency would change the behavior of $\chi$ under rotations, since the transformation of the $\mathbf{q}$-modulated sums over lattice vectors arising from Eq.~(\ref{eq:qDependentEffectiveR}) will have to be taken into account.

On a basis of spherical tensors, the matrix $D$ can be expressed using Wigner-D symbols. Since rotations conserve angular momentum, the transformation matrix $D$ is block diagonal, structured as:
\begin{align}
D(\phi,\theta,\psi) = \left(
    \begin{tabular}{ccccccccc}
        $D^{(0)}$ & 0 & 0 & 0 & 0 & 0 & 0 & 0 & 0 \\
        0 &  &  &  & 0 & 0 & 0 & 0 & 0 \\
        0 &  \multicolumn{3}{c}{$D^{(1)}$}  & 0 & 0 & 0 & 0 & 0 \\
        0 &  &  &  & 0 & 0 & 0 & 0 & 0 \\
        0 & 0 & 0 & 0 &  &  &  &  &  \\
        0 & 0 & 0 & 0 &  &  &  &  &  \\
        0 & 0 & 0 & 0 &  \multicolumn{5}{c}{$D^{(2)}$}  \\
        0 & 0 & 0 & 0 &  &  &  &  &  \\
        0 & 0 & 0 & 0 &  &  &  &  &  
    \end{tabular}
    \right),
\end{align}
where $D^{(l)}(\phi,\theta,\psi)$ is a Wigner-D symbol that describes the rotation of a spherical tensor of rank $l$. For a matrix expressed in a basis of spherical tensors, the spherical average is given by the trace of each angular momentum sector. Thus, we obtain:
\begin{align}
\overline{\chi} = \left(
    \begin{tabular}{ccccccccc}
        $\overline{\chi}^{(0)} \mathds{1} $ & 0 & 0 & 0 & 0 & 0 & 0 & 0 & 0 \\
        0 &  &  &  & 0 & 0 & 0 & 0 & 0 \\
        0 &  \multicolumn{3}{c}{$\overline{\chi}^{(1)} \mathds{1} $}  & 0 & 0 & 0 & 0 & 0 \\
        0 &  &  &  & 0 & 0 & 0 & 0 & 0 \\
        0 & 0 & 0 & 0 &  &  &  &  &  \\
        0 & 0 & 0 & 0 &  &  &  &  &  \\
        0 & 0 & 0 & 0 &  \multicolumn{5}{c}{$\overline{\chi}^{(2)} \mathds{1}$}  \\
        0 & 0 & 0 & 0 &  &  &  &  &  \\
        0 & 0 & 0 & 0 &  &  &  &  &  
    \end{tabular}
    \right),
    \label{eq:chiOverline}
\end{align}
where $\overline{\chi}^{(l)}$ is given by $1/(2l+1)$ times the trace over the corresponding sub-block of the full RIXS tensor $\chi$:
\begin{align}
    \overline{\chi}^{(0)} &= \chi_{s,s}, \\ \nonumber
    \overline{\chi}^{(1)} &= \frac{1}{3}\big( \chi_{R_x,R_x} + \chi_{R_y,R_y} + \chi_{R_z,R_z} \big), \\ \nonumber
    \overline{\chi}^{(2)} &= \frac{1}{5}\big( \chi_{d_{x^2-y^2},d_{x^2-y^2}} + \chi_{d_{z^2},d_{z^2}} \\ \nonumber 
    & \quad\quad\quad + \chi_{d_{yz},d_{yz}} + \chi_{d_{xz},d_{xz}} + \chi_{d_{xy},d_{xy}} \big).
    \label{eq:chiOverlineComponents}
\end{align}

For an experiment on a powder sample where the outgoing light polarization is not analyzed (Eq.~(\ref{eq:nopolfilterout})), the RIXS tensor simplifies to:
\begin{align}
    \overline{\chi}_{\boldsymbol{\hat{\pi}}_{\mathrm{in}}} &= \frac{1}{6}\left(\overline{\chi}^{(0)} + 2 \overline{\chi}^{(2)}\right)\cos^2(2\theta) \\ \nonumber 
    & \quad + \frac{1}{4}\left(\overline{\chi}^{(1)} +\overline{\chi}^{(2)} \right) \left( 1 + \sin^2(2\theta) \right), \\ \nonumber
    \overline{\chi}_{\boldsymbol{\hat{\sigma}}_{\mathrm{in}}} &= \frac{2}{12} \overline{\chi}^{(0)} + \frac{3}{12} \overline{\chi}^{(1)} + \frac{7}{12} \overline{\chi}^{(2)}.
\label{eq:angleDependetScattering}
\end{align}

For the considered case of core-to-core RIXS, where the transferred momentum does not play a significant role and the scattering angle $2\theta$ can be varied, these results demonstrate that three distinct fundamental spectra can be measured as a function of the incoming light polarization. Notably, no polarization filter for the scattered light is required.

For core-to-valence RIXS, however, Eqs.~(\ref{eq:ChiIntegralAvg}) and (\ref{eq:chiOverline}) are generally not valid and the transferred momentum does influence the results. In this case, varying the scattering angle $2\theta$ also changes the transferred momentum, requiring additional considerations in the analysis.

\section{\label{sec:Analyzer}The effect of the analyzer}

\begin{figure}[htbp]
    \includegraphics[width=0.7\linewidth]{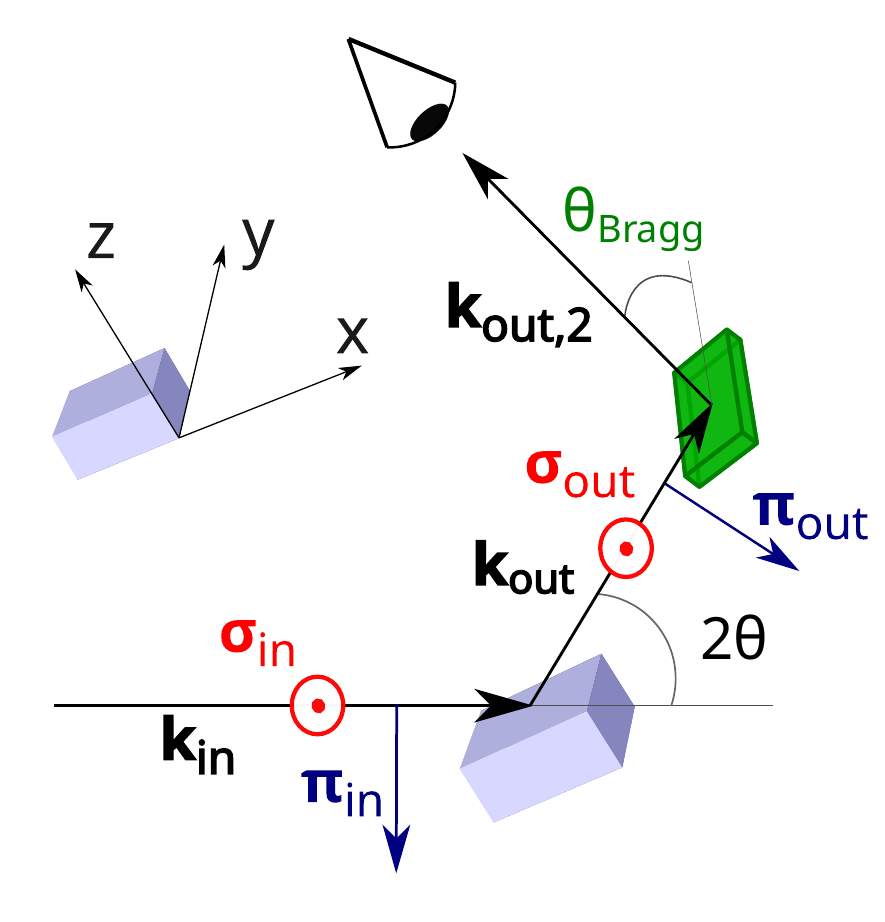}
    \caption{\label{fig:FigScatteringGeometrydetector} Schematic representation of a general RIXS geometry with scattering angle $2\theta$. After RIXS, the outgoing photons from the sample elastically scatter at a Bragg crystal or grating (green) before being detected. The RIXS scattering plane can differ from the Bragg plane. $\hbar\mathbf{k}$ represents the photon momentum. $\bm{\mathrm{\sigma}}$ and $\bm{\mathrm{\pi}}$ are the vertical (out-of-plane) and horizontal (in-plane) polarization basis vectors, respectively.}
\end{figure}

In the previous sections, we discussed how the measured RIXS spectra depend on the polarization of the incoming and scattered light. There is, however, one more crucial aspect to consider when predicting and interpreting the intensities observed in real experiments. In Fig.~\ref{fig:FigScatteringGeometrySample}, we outlined the general experimental geometry. We sofar described the polarization-dependent intensity of the scattered light with wave vector $\mathbf{k}_{\mathrm{out}}$. 

To measure the photon-energy-dependent intensity of this scattered light, a detector is required. Most experimental setups include a grating or analyzer crystal to filter the scattered photons by energy. Figure~\ref{fig:FigScatteringGeometrydetector} shows the experimental setup, now including an analyzer grating or crystal that reflects the scattered light via Bragg diffraction. Importantly, this Bragg reflection does not necessarily occur within the RIXS scattering plane defined by $\mathbf{k}_{\mathrm{in}}$ and $\mathbf{k}_{\mathrm{out}}$. 

Beyond filtering by energy, the analyzer crystal also acts as a polarization filter. While it does not rotate the polarization, the Bragg reflection ensures that the polarization of the transmitted light must remain perpendicular to its new Poynting vector, which is parallel to the wave vector $\mathbf{k}_{\mathrm{out},2}$ of the scattered photons after the analyzer.

In the previous sections, we used the unit polarization basis vectors $\boldsymbol{\hat{\sigma}}_{\mathrm{out}}$ and $\boldsymbol{\hat{\pi}}_{\mathrm{out}}$, defining the polarization of the scattered light $\boldsymbol{\hat{\epsilon}}_{\mathrm{out}}$ accordingly. After passing through the analyzer crystal, the observed intensity is modified, and we define an effective polarization vector that is no longer normalized:
\begin{align}
    \boldsymbol{{\epsilon}}_{\mathrm{out}}^{\,\,\mathrm{eff}} = \left(1-\left|\boldsymbol{\hat{\epsilon}}_{\mathrm{out}} \cdot \mathbf{\hat{k}}_{\mathrm{out},2}\right|\right) \boldsymbol{\hat{\epsilon}}_{\mathrm{out}}.
\end{align}

This effective polarization vector can be used in all previous equations in place of $\boldsymbol{\epsilon}_{\mathrm{out}}$ to correctly determine the detected scattered intensity. Instead of considering the full effective polarization vector, it is often useful to analyze its $\sigma$ and $\pi$ components separately. These components are defined as:
\begin{align}
        \boldsymbol{\sigma}_{\mathrm{out}}^{\,\,\mathrm{eff}} = \left(1-\left|\boldsymbol{\hat{\sigma}}_{\mathrm{out}} \cdot \mathbf{\hat{k}}_{\mathrm{out},2}\right|\right) \boldsymbol{\hat{\sigma}}_{\mathrm{out}},
\end{align}
\begin{align}
        \boldsymbol{\pi}_{\mathrm{out}}^{\,\,\mathrm{eff}} = \left(1-\left|\boldsymbol{\hat{\pi}}_{\mathrm{out}} \cdot \mathbf{\hat{k}}_{\mathrm{out},2}\right|\right) \boldsymbol{\hat{\pi}}_{\mathrm{out}}.
\end{align}

These effective polarization vectors can now be used in the equations from the previous sections, in particular Eq.~(\ref{eq:nopolfilterout}), to account for the effect of the Bragg reflection on the polarization-dependent intensity.

\section{Conclusion}
In this work, we demonstrated how the polarization dependence of the scattered light in RIXS can be described using a RIXS tensor $\chi$. This cartesian is a rank-4 tensor. To systematically analyze its components in systems with symmetry, we coupled the light polarization vectors $\boldsymbol{\hat{\epsilon}}_{\mathrm{in}}$ and $\boldsymbol{\hat{\epsilon}}_{\mathrm{out}}$ to form a new vector, $\hat{\bf{e}} = \{\boldsymbol{\hat{\epsilon}}_{\mathrm{in}} \otimes \boldsymbol{\hat{\epsilon}}_{\mathrm{out}}^*\}$, with nine components. Expressing this vector in a spherical tensor basis with angular momentum components $l=0$, $l=1 (R)$, and $l=2$ provided a convenient framework to highlight the symmetry properties of $\chi$. In this representation, only a limited number of tensor elements—those associated with the irreducible representations of the point group of the sample —remain nonzero. 

Understanding the polarization dependency in RIXS also helps to determine the character of observed excitations. This has been experimentally done, for example, to distinguish between charge- and spin-like excitations in 2p3d RIXS of Antiferromagnetic Nd$_2$CuO$_4$ \cite{kang_resolving_2019}. The introduced framework allows to expand the concept into a more general approach. Here, each fundamental spectrum with character $\Gamma_{\chi}$ couples the initial state belonging to $\Gamma_{g}$ to final states with characters $\Gamma_{f}\in \{\Gamma_{g}\otimes\Gamma_{\chi}\}$. For the special cases where the ground state transforms as the trivial representation ($\Gamma_{g}=a_{g}$) then $\Gamma_{f}=\Gamma_{\chi}$ is a strict selection rule. This is the case in 3d4f RIXS of U$^{VI}(5f^0)$, where the final state satellite has total J=0 ($\Gamma_{f}=a_g$) \cite{schacherl_resonant_2025}. At $2\theta=\pi/2$, following Eqs.~(\ref{eq:chiOverlineComponents}) and (\ref{eq:angleDependetScattering}), this feature can be reached only by $\overline{\chi}^{(0)}$ and thus visible in $\overline{\chi}_{\boldsymbol{\hat{\sigma}}_{\mathrm{in}}}$ but not in $\overline{\chi}_{\boldsymbol{\hat{\pi}}_{\mathrm{in}}}$. 

In summary, the notation introduced in this work allows for a clear separation between the experimental geometry and the intrinsic material properties of the studied system. This approach enables a straightforward comparison of expected intensities across different measurement configurations, including  powder samples and experiments conducted with or without polarization analysis of the scattered light. Extensions of the presented results to include momentum dependent scattering is possible and will be focus of future investigations.

This work is supported by the Deutsche Forschungsgemeinschaft (DFG, German Research Foundation) through the research unit QUAST, FOR 5249 (Project P7), Project ID No. 449872909. The authors acknowledge support by the state of Baden-Württemberg through bwHPC and the German Research Foundation (DFG) through grant no INST 40/575-1 FUGG (JUSTUS 2 cluster). 

\bibliography{referencesTensor}
\end{document}